\documentclass[reprint,english,aps,prl,twocolumn,amsmath, amssymb, showpacs]{revtex4}
\usepackage[T1]{fontenc}
\usepackage{subfigure}
\usepackage[latin1]{inputenc}
\usepackage{graphicx}
\usepackage{epstopdf}
\usepackage{epsfig}
\usepackage{prettyref}
\usepackage{babel}
\usepackage{color}
\makeatother

\begin{document}

\title{Tuning the phase transition dynamics by variation of cooling field and metastable phase fraction in Al doped Pr$_{0.5}$Ca$_{0.5}$MnO$_3$}
%\title{Weak ferromagnetism and spin-glass like characteristics in the magnetically ordered state of the perovskite oxide NdNiO$_3$}

\author{Devendra Kumar}
\affiliation{UGC-DAE Consortium for Scientific Research, University Campus, Khandwa Road, Indore-452001,India}

\author{Kranti Kumar}
\affiliation{UGC-DAE Consortium for Scientific Research, University Campus, Khandwa Road, Indore-452001,India}

\author{A. Banerjee}
\affiliation{UGC-DAE Consortium for Scientific Research, University Campus, Khandwa Road, Indore-452001,India}

\author{P. Chaddah}
\affiliation{UGC-DAE Consortium for Scientific Research, University Campus, Khandwa Road, Indore-452001,India}

\begin{abstract}
We report the effect of field, temperature and thermal history on the time dependence in resistivity and magnetization in the phase separated state of Al doped Pr$_{0.5}$Ca$_{0.5}$MnO$_3$. The rate of time dependence in resistivity is much higher than that of magnetization and it exhibits a different cooling field dependence due to percolation effects. Our analysis show that the time dependence in physical properties depends on the phase transition dynamics which can be effectively tuned by variation of temperature, cooling field and metastable phase fraction. The phase transition dynamics can be broadly divided into the arrested and un-arrested regimes, and in the arrested regime, this dynamics is mainly determined by time taken in the growth of critical nuclei. An increase in cooling field and/or temperature shifts this dynamics from arrested to un-arrested  regime, and in this regime, this dynamics is determined by thermodynamically allowed rate of formation of critical nuclei which in turn depends on the cooling field and available metastable phase fraction. At a given temperature, a decrease in metastable phase fraction shifts the crossover from arrested to un-arrested regimes towards lower cooling field. It is rather significant that  inspite of the metastable phase fraction calculated from resistivity being somewhat off from that of magnetization,  their cooling field dependence exhibits a striking similarity which indicate that the dynamics in arrested and un-arrested regimes are so different that it comes out vividly provided that the measurements are done around percolation threshold.
%We report the effect of field, temperature and thermal history on the dynamics of physical properties in phase separated state of Al doped Pr$_{0.5}$Ca$_{0.5}$MnO$_3$. Our analysis show that these dynamical effects depend on the phase transition dynamics which can be effectively
%tuned by variation of temperature, cooling field and metastable phase fraction. The phase transition dynamics can be broadly
%divided in the arrested and un-arrested regimes, and in the arrested regime, this dynamics is mainly determined by time taken in atomic
%rearrangement. An increase in cooling field and temperature shifts this dynamics from arrested to un-arrested  regime, and in this
%regime, this dynamics is determined by thermodynamically allowed rate of transformation which in turn depends on the cooling field and available
%metastable phase fraction. At a given temperature, a decrease in metastable phase fraction shifts the crossover from arrested to un-arrested dynamics
%towards the lower cooling field.
\end{abstract}

\pacs{75.47.Lx , 75.30.Kz , 64.60.My}

%\keywords{Phase-Transition Dynamics, Nucleation, Percolation, Kinetic Arrest}

\maketitle

\section{Introduction}
The study of broad first order phase transition and the associated phase  separated (PS) state in the transition metal oxides has drawn a significant attention in the last few decades.\cite{Dagotto} The PS state consists of coexisting metastable and stable phases with distinct free energies, and in this state the physical properties of the system can be varied dramatically on changing the fields such as magnetic field, pressure, electric field etc.\cite{Kuwahara, Morimoto, Asamitsu} These interesting manifestations of the phase coexistence are mainly attributed to field induced variation in phase fraction of the coexisting phases.\cite{Banerjee4} A large number of such PS systems exhibit dynamical effects, such as, time dependence in physical properties,\cite{Granados, Chen, Levy, Chaddah11, Ghivelder, Banerjee, Banerjee1, Devendra, Devendra1, Devendra2} cooling and  heating rate dependence,\cite{Uehara1,Levy,Devendra} 1/f noise,\cite{Podzorov} and aging and rejuvenation effects.\cite{Levy, Devendra} The existence of these dynamical effects in PS state has been attributed to the transformation of the metastable phases, each having a distinct temperature of metastability, to stable state.\cite{Chaddah11, Devendra, Devendra1, Devendra2} This transformation is a two step process, the first step is the formation of critical nuclei and its rate as well as size of formation is decided by the difference in free energy of metastable and stable state, while the second step is the growth of this critical nuclei to complete the phase transformation and the rate of growth %which is decided by the diffusion rate that in turn
depends on the temperature of measurement.\cite{Lifshitz, Callister}  %In the case of manganites, the time constant associated with the structural interface motion (during the growth of critical nuclei) is of the order of micro-seconds.\cite{Sharma} This suggests that the time taken in the growth of critical nuclei is small and so the rate of phase transformation is mainly determined by the time taken in the formation  of critical nuclei. % and the allowed rate of transformation is found to depend on the energy of nucleation.\cite{Devendra, Devendra2}
%The existence of these dynamical effects in PS state has been attributed to the transformation of the metastable phases, each having a distinct temperature of metastability, to stable state and the allowed rate of transformation is found to depend on the energy of nucleation.\cite{Devendra, Devendra2}

The growth of stable phase critical nuclei in the metastable phase happens through the motion of stable-metastable phase interface which in turn depends on the kinetics of constituent atoms, and the recent resonant ultrasound spectroscopy measurements on  La$_{0.215}$Pr$_{0.41}$Ca$_{3/8}$MnO$_3$ suggest that the time constant associated with this interface motion (or growth of critical nuclei) is of the order of micro-seconds.\cite{Sharma} On lowering the temperature  atomic motion slows down, and so, some of the metastable phases may approach their respective temperature of kinetic arrest ($T_g$). Below $T_g$, the dynamics of stable-metastable phase interface motion freezes which blocks the growth of critical nuclei.\cite{Sharma} The freezing of interface motion hinders the metastable to ground state transformation at experimental timescale or in other words the kinetics of the first order transformation is arrested below $T_g$.\cite{Debenedetti} For a magnetic transition, the experiments performed on a wide range of materials show that the temperature of kinetic arrest $T_g$ depends on the applied field and for ferromagnetic ground state $T_g$ decreases on increasing the field while for antiferromagnetic ground state $T_g$ increases on increasing the field.\cite{Banerjee,Kumar,Banerjee1} This result has been argued on heuristic ground and is found to be consistent with the  Le Chatelier's principle.\cite{Banerjee4, Lakhani2} In these systems, the kinetics  of a phase can change from arrested to un-arrested regime on varying the magnetic field and temperature. To understand the dynamical effects observed in the phase-separated state it would be quite intriguing to investigate  how the nucleation and growth process are effected by the magnetic field and temperature and how their field and temperature dependence affect the dynamical properties.%   Therefore it is quite intriguing to examine the  field dependence of these two distinct dynamical process and how their field dependence affect the observed dynamical properties.

To address these issues we have performed detailed field, temperature and thermal history dependent time decay measurements on 2.5\% Al doped
Pr$_{0.5}$Ca$_{0.5}$MnO$_3$ (PCMAO). Most of the functional applications of such perovskite oxides use their magneto-resistive properties in the phase-separated state.\cite{Uehara} This is because the metallic and insulating phases of these systems have a wide difference in resistivity, and in the percolation regime,
a small change in phase fraction on varying the field or temperature manifest in a  huge change in resistivity. Therefore in the percolative regime
of the phase separation, the resistivity is very sensitive to the change in phase fraction, and accordingly we have carried out time dependent resistive measurements
in the percolative regime to observe the optimum effect of phase transition dynamics. The conclusions drawn from the resistivity measurements are also corroborated
by magnetization measurements. PCMAO is antiferromagnetic insulator (AF-I) at room temperature and has a  ferromagnetic metallic (FM-M) ground state, but
on zero field cooling the kinetics of AF-I to FM-M transition gets arrested and therefore PCMAO remains in AF-I state even at low temperatures.\cite{Banerjee1}
On increasing the cooling field, the kinetics of a fraction of AF-I phases does not arrest during cooling, and the transformation of these un-arrested AF-I
phases to FM-M state enhances the FM-M phase fraction at low temperatures.\cite{Banerjee3, Lakhani2}

%To address these issues we have performed detailed field, temperature and thermal history dependent time decay measurements on 2.5\% Al doped  Pr$_{0.5}$Ca$_{0.5}$MnO$_3$ (PCMAO). PCMAO is antiferromagnetic insulator (AF-I) at room temperature and has a  ferromagnetic metallic (FM-M) ground state, but on zero field cooling the kinetics of AF-I to FM-M transition gets arrested and therefore PCMAO remains in AF-I state even at low temperatures.\cite{Banerjee1} On increasing the cooling field, the kinetics of a fraction of AF-I phases does not arrest during cooling, and the transformation of these un-arrested AF-I phases to FM-M state enhances the FM-M phase fraction at low temperatures.\cite{Banerjee3, Lakhani}

 % On increasing the cooling field, $T_g$  decreases and so the kinetics of a fraction of AF-I phases gets de-arrested. The transformation of these de-arrested AF-I phases to FM-M state during cooling enhances the FM-M phase fraction at low temperatures.\cite{Banerjee3, Lakhani} %these AF-I phases will transform to FM-M state in the process of cooling. %A further increase in cooling field enhances the dearrested AF-I  phase fraction and the transformation of these AF-I phases to FM-M state enhances the FM-M phase fraction at low temperatures.
%Below AF-I to FM-M transition temperature, an AF-I phase may exists in its metastable state and the transformation of such AF-I phases to FM-M state can give rise to time dependence in physical properties.

 In this work we have performed detailed temperature, time  and cooling field dependent resistivity and magnetization measurements on PCMAO to understand the effect of various dynamic mechanisms on the rate of phase transition and the associated dynamical effects. Our results show that the time dependent effects observed in physical properties of PCMAO depends on cooling field, temperature of measurement, and thermal cycling. At a fixed temperature, an increase in cooling field shifts the kinetics of the metastable AF-I phases from arrested to un-arrested regime. Further, this crossover from arrested to un-arrested dynamics happens at a lower cooling field on reducing the non-equilibrium phase fraction by thermal cycling (heating run).  A detailed analysis of the experimental results show that in the un-arrested regime, the observed rate of transformation from metastable AF-I to FM-M state is determined by thermal energy, energy of nucleation, and volume fraction of metastable AF-I phases. At constant temperature, an increase in magnetic field reduces the energy of nucleation and the volume fraction of metastable AF-I phases, and so the field dependence of the observed rate of transformation is decided by the competition between the field dependence of these two factors.
\section{Experimental Details}
The Pr$_{0.5}$Ca$_{0.5}$Mn$_{0.975}$Al$_{0.025}$O$_3$ (PCMAO) sample used in this study is the same as used in Refs. \cite{Banerjee1, Banerjee3, Lakhani2, Nair1}. The polycrystalline samples of PCMAO are prepared by standard solid state reaction technique using 99.99~\% pure Pr$_6$O$_{11}$, CaCO$_3$, MnO$_2$, and Al$_2$O$_3$ in stoichiometric ratio. The details of sample preparation and characterization are given elsewhere.\cite{Nair}  The resistivity and magnetization measurements were performed on Quantum Design 14 Tesla physical properties measurement system-vibrating sample magnetometer system. The magnetic field is changed only at room temperature and the super-conducting  magnet was kept in persistent mode during cooling, heating, and the time decay measurements. The cooling and heating rate of 1.5~K/min is used throughout the measurements.
\section{Results and Discussion}

% figure 1
\begin{figure}[!t]
\begin{centering}
\includegraphics[width=1\columnwidth]{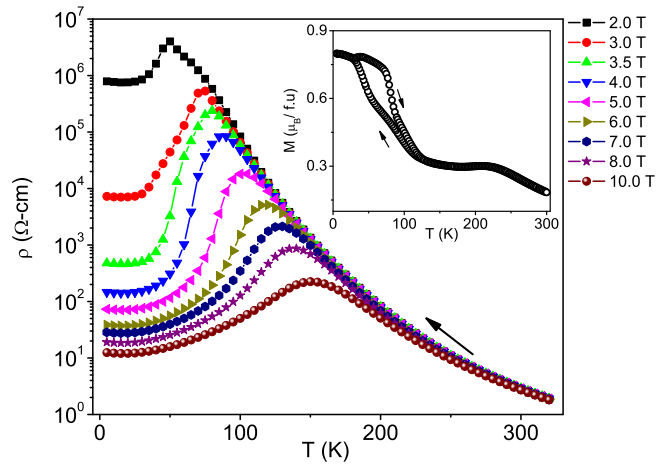}
\par\end{centering}
\caption{Resistivity versus temperature plot for PCMAO in the cooling run at cooling fields of 2-10~T. The inset show the magnetization versus temperature cycle at the measuring field of 3~T.} \label{fig: RvsT}
\end{figure}

Figure \ref{fig: RvsT} show the resistivity of  PCMAO  in cooling run in various magnetic fields. We can see that PCMAO which remains an insulator in absence of field, exhibits a broad insulator to metal transition at low temperatures on cooling in presence of 2~T or higher field. The inset of figure \ref{fig: RvsT} display the magnetization of PCMAO at 3~T field. The magnetization also exhibits a concurrent antiferromagnetic (AF) to ferromagnetic (FM) transition with a thermal irreversibility in the cooling and heating cycle which confirms that the transition from antiferromagnetic insulating (AF-I) to ferromagnetic metallic (FM-M) state is a disordered broadened first-order transition. The resistivity curves show that on increasing the cooling field, the AF-I to FM-M transition shifts towards higher temperature with a simultaneous drop of the low temperature resistivity. The relative drop in low temperature resistivity is huge at lower cooling fields ($\sim$2T) and it decreases on increasing the cooling field (see figure \ref{fig: RvsT}). This drop in low temperature resistivity is due to  field induced variation of the phase fraction of coexisting FM-M and AF-I phases which determine the resistivity in association with percolation effects. % and the observed cooling field dependence indicates that possibly an increase in cooling field moves the system away from the regime of percolative transport.
Our earlier work on PCMAO has established that the PCMAO has FM-M ground state and the fraction of AF-I phase present is the outcome of the glass like arrest of the kinetics of first order AF-I to FM-M phase transition. The fraction of arrested AF-I phases can be tuned by choosing different path in H-T space.\cite{Banerjee1, Banerjee3, Lakhani2} In the cooling process, resistivity deviates from the insulating curve when the FM-M phase starts appearing, and on further lowering the temperature the FM-M phase fraction increases and resistivity changes sign of the slope. Cooling below this temperature, the FM-M phase fraction further increases causing a further drop in resistivity, and around 28~K, slope of the resistivity becomes nearly constant indicating the slow down of kinetics of AF-I to FM-M phase transformation. This change in the slope of resistivity around 28~K is steep at low fields and it becomes gradual on increasing the field.

\begin{figure} [!t]
\begin{centering}
\includegraphics[width=1.0\columnwidth]{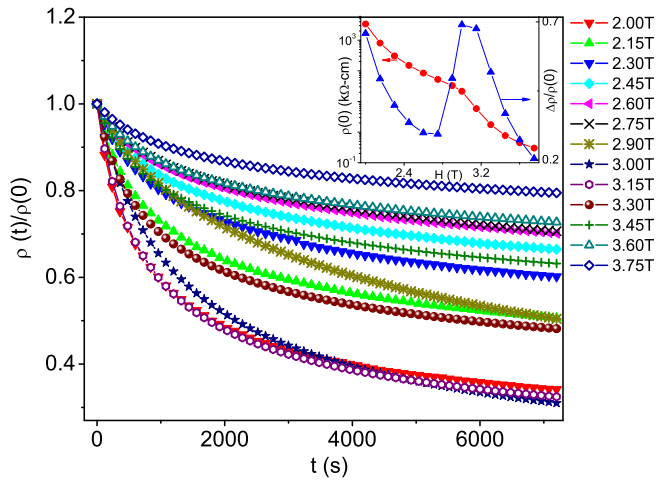}
\par\end{centering}
\caption{Time dependence of resistivity at 28~K in the cooling run at cooling fields of 2-3.75~T. The inset show the $\rho(0)$ (resistivity at t=0) and $\Delta\rho/\rho(0)$ (the normalized change in resistivity in two hours) for different cooling fields.} \label{fig:restime28K}
\end{figure}

To understand the dynamics of transformation of non-equilibrium AF-I phase into equilibrium FM-M state, particularly to study the effect of control parameters such as field, temperature, and path in H-T space on the approach towards equilibrium, we have performed time dependence measurements at 28~K, 25~K and 24~K using different path in H-T space.  %From figure \ref{fig: RvsT} and its inset, it is clear that in the phase separated state, the resistivity is more sensitive
%to the change in phase fraction in comparison to the magnetization. Moreover, the resistivity can be determined more accurately than the magnetization.
%Keeping these points in mind, we have used resistivity as a primary probe for the time dependence measurements.
In the cooling cycle measurements, we cool the sample from room temperature to temperature of interest in presence of field and then resistivity is recorded as a function of time keeping field constant. Figure \ref{fig:restime28K} show the time dependent resistivity data at 28~K in cooling run for various cooling fields. The data is presented as $\rho(t)/\rho(0)$ (where $\rho(0)$ is the resistivity at t=0) so that the different field values are normalized for easy comparison. We note that the resistivity decreases with time which indicates that the AF-I phases, which are in their supercooled metastable state, slowly transform to the FM-M ground state. The inset of figure \ref{fig:restime28K} show the cooling field dependence of $\rho(0)$ and $\Delta\rho/\rho(0)$ (which is the the normalized change in resistivity in two hours). The $\rho(0)$ decreases on increasing the cooling field and its field dependence show a change in behavior at around 2.9~T.  This change in cooling field dependence of $\rho(0)$ suggests that possibly the percolation threshold of FM-M phases occurs around 2.9~T.  The $\Delta\rho/\rho(0)$ gives the rate of resistivity decay; it decreases on increasing the field and attains a minimum around 2.75~T (close to the point of percolation threshold of FM-M phases). On further increasing the field, the rate of resistivity decay grows and attains a maximum around 3.15~T and thereafter it decreases.
%The time dependent resistivity data fits well with the stretched exponential relaxation function
%\begin{equation}
%\rho(t)=\rho(0)-\rho(1)(1-e^{-\left(\frac{t}{\tau}\right)^{\gamma}})\label{eq:Stretched-Exponential1}
%\end{equation}
%where the $\rho(1)$, $\tau$, and $\gamma$ are fitting coefficients. The variation of $\rho(0)$ and $\rho(1)/\rho(0)$ with field are shown in inset of figure \ref{fig:restime28K}. The $\rho(0)$ decreases on increasing the cooling field and it shows a kink at around 2.9~T which is possibility due to percolation of FMM phases.  The $\rho(1)/\rho(0)$ gives the rate of resistivity decay; it decreases on increasing the field and attains a minimum around 2.75~T (close to the point of percolation threshold of FMM phases). On further increasing the field the rate of resistivity decay grows and attains a maximum around 3.15~T and thereafter it decreases on increasing the field.
%figure 2

\begin{figure} [!t]
\begin{centering}
\includegraphics[width=1.0\columnwidth]{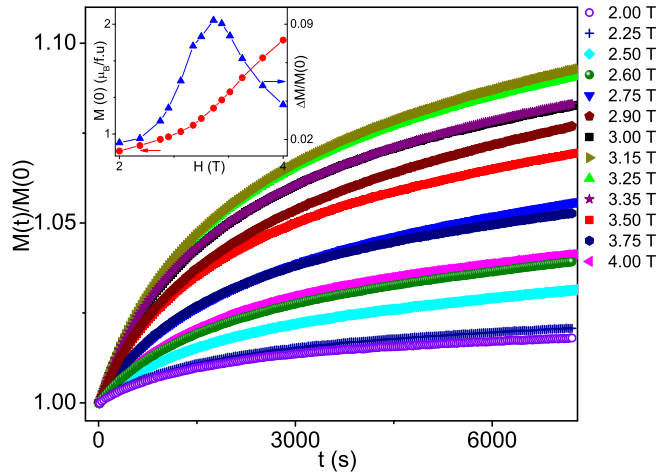}
\par\end{centering}
\caption{Time dependence of magnetization at 28~K in the cooling run at different cooling fields. The field was kept constant throughout the measurement. The inset show the field variation of M(0) (magnetization at t=0) and $\Delta$M/M(0) (the normalized change in magnetization in two hours).} \label{fig: timdepMag}
\end{figure}

To comprehend the resistivity measurements, we have also performed time dependent magnetization measurements at 28~K. In these measurements we cool the system from room temperature to temperature of interest in presence of field, wait for temperature stabilization and then record magnetization as a function of time keeping field constant. The time dependence of magnetization at 28~K for different cooling fields is shown in figure \ref{fig: timdepMag}. The data is presented as $M(t)/M(0)$ so that different field data is normalized for easy comparison. The magnetization increases with time which shows that FM-M phase fraction increases at the expense of AF-I phase. The inset of figure \ref{fig: timdepMag} show the variation of M(0) (the magnetization at t=0 for different fields) and $\Delta$M/M(0) (change in magnetization in 2 hours normalized with M(0)). The $\Delta$M/M(0) represents the growth rate of magnetization; it increases on increasing the field, attains a maximum at 3.15~T, and thereafter decreases on further increasing the field.

The time dependence in resistivity and magnetization are due to same physical mechanism namely the transformation of AF-I metastable phases to FM-M stable state, and so, their rate of change $\Delta\rho/\rho(0)$ and $\Delta$M/M(0) should exhibit similar field dependence. But from the inset of figures \ref{fig:restime28K} and \ref{fig: timdepMag} we can see that $\Delta\rho/\rho(0)$ is around an order of magnitude larger than $\Delta$M/M(0) and it exhibits a different field dependence than that of $\Delta$M/M(0). While $\Delta\rho/\rho(0)$ and $\Delta$M/M(0) both exhibit a peak around 3.15~T, the $\Delta\rho/\rho(0)$ show an extra minima around 2.75~T which is absent in $\Delta$M/M(0). Since magnetization is a more direct probe of phase fractions, the larger magnitude of $\Delta\rho/\rho(0)$ and the presence of extra minima in $\Delta\rho/\rho(0)$ versus H curve suggest that possibly these extra feature in resistivity are coming from percolation effects.  %Since magnetization is a more direct probe of phase fractions, the presence of extra minima in the $\Delta\rho/\rho(0)$ versus H curve suggest that possibly this extra feature in resistivity is coming from percolation effects.

\begin{figure} [!t]
\begin{centering}
\includegraphics[width=1.0\columnwidth]{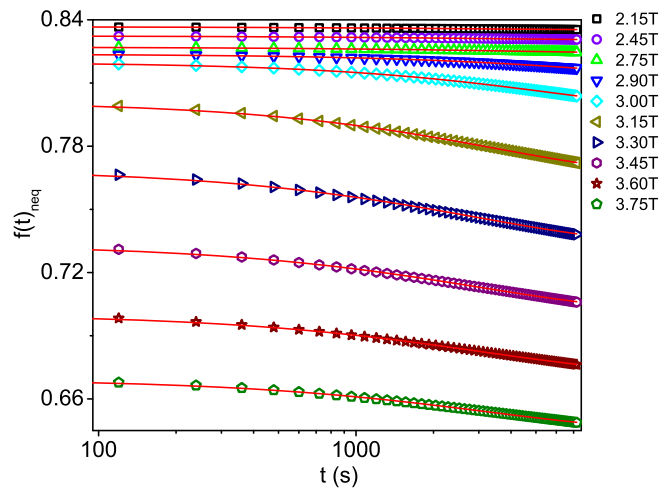}
%instead of fig15.eps
\par\end{centering}
\caption{Time dependence of non equilibrium volume fraction ($f_{neq}$) at 28~K in cooling run at cooling fields between 2 to 3.75~T. The $f_{neq}$  is calculated using the time dependent resistivity data of figure \ref{fig:restime28K} and the GEM equation. Not all the data are shown here to avoid clutter. The red lines are the fitting of equation \ref{eq:Stretched-Exponential} to $f(t)_{neq}$.} \label{fig:timedepvolume}
\end{figure}

\begin{figure} [!t]
\begin{centering}
\includegraphics[width=1.0\columnwidth]{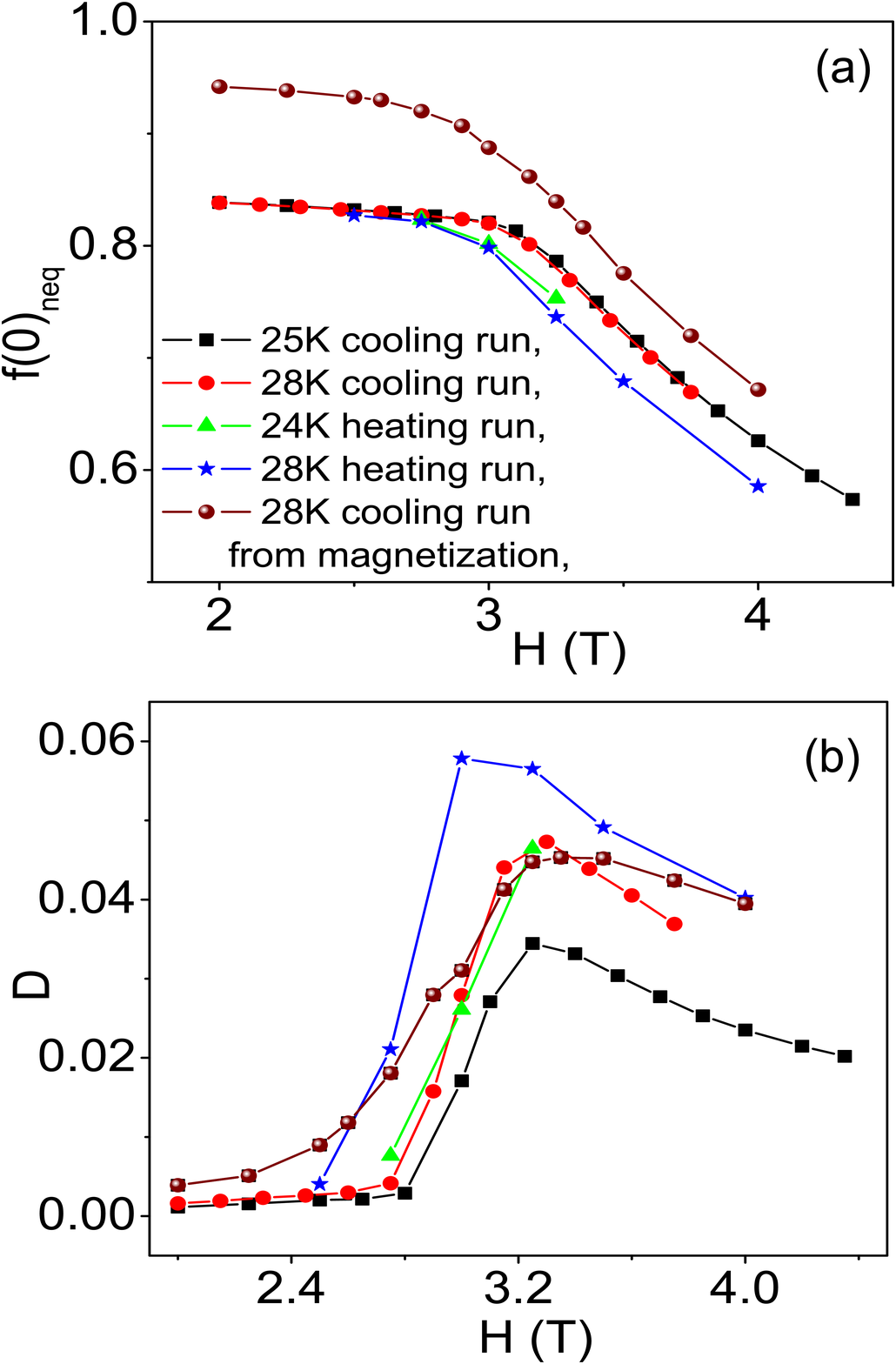}
%instead of fig15.eps
\par\end{centering}
\caption{(a) $f(0)_{neq}$ and (b) $D$  versus cooling field at different temperatures and thermal cycle. The symbols in figure (a) and (b) are the same for a particular temperature and run.} \label{fig: Decayrate}
\end{figure}

To investigate it further we note that the resistivity of PCMAO is coupled to AF-I and FM-M phase fraction through percolation effects, and so, the decay rate of resistivity does not exactly represent the time variation of phase fractions. To get an idea of the time dependence of phase fractions, we have converted the time dependent resistivity data into the phase fraction using McLachlan formula, which is based on a general effective medium (GEM) theory.\cite{McLachlan} If $\sigma_{E}$ is the effective electrical conductivity  of a binary MI mixture, the GEM equation says
\begin{equation}
(1-f)\frac{(\sigma_{I}^{1/t}-\sigma_{E}^{1/t})}{(\sigma_{I}^{1/t}+A\sigma_{E}^{1/t})}+f\frac{(\sigma_{M}^{1/t}-\sigma_{E}^{1/t})}{(\sigma_{M}^{1/t}+A\sigma_{E}^{1/t})}=0\label{eq:GEM}
\end{equation}
where $f$ is the volume fraction of the metallic phases and $A=(1-f_{c})/f_{c}$, $f_{c}$ being the volume fraction of metallic phases at the percolation threshold, and $t$ is a critical exponent which is close to 2 in three dimensions.\cite{Herrmann,Hurvits}  The constant $f_{c}$ depends on the lattice dimensionality, and for 3D its value is 0.16.\cite{Efros} The GEM equation has been successfully applied to a wide variety of isotropic, binary, macroscopic mixtures and its applicability has been also tested on phase separated manganite La$_{5/8-x}$Pr$_x$Ca$_{3/8}$MnO$_3$ where it gives satisfactory results with standard  $t$, but at very low temperatures $t$=4 gives better results.\cite{Kim} In order to calculate the volume fraction of AF-I and FM-M phases, we need the resistivity of AF-I and FM-M metallic phases. Since below 28~K the change in resistivity on going from 8~T to 10~T is quite small, we can assume that at 10~T cooling field the system is almost in FM-M state.\cite{endnote1} With this assumption, at 28~K or below, the FM-M phase resistivity can be obtained from 10~T cooled resistivity data. The AF-I phase resistivity is calculated by extrapolating the resistivity curves above the transition to low temperatures. Using these values, we have converted the time dependent resistivity data into the time dependent AF-I (non-equilibrium) phase fraction $f_{neq}$. A subset of time dependent resistivity data converted to non-equilibrium phase fraction is shown in figure \ref{fig:timedepvolume}. The time decay of non-equilibrium phase fraction fits best with the stretched exponential relaxation function
\begin{equation}
f_{neq}(t)=f_{neq}(0)(1-D(1-e^{-\left(\frac{t}{\tau}\right)^{\gamma}})\label{eq:Stretched-Exponential}
\end{equation}
where $f_{neq}(0)$ is the phase fraction of non-equilibrium phase at t=0; $D$, $\tau$, and $\gamma$ are fit parameters. The parameter $D$ is the rate constant that determines how fast conversion from AF-I to FM-M phase will occur and what fraction of volume will finally convert. In figure \ref{fig: Decayrate} we have plotted the cooling field dependence of $f_{neq}(0)$ and $D$  and we can see that the extra minimum that is present in the decay rate of resistivity disappears when the decay rate of resistivity is converted to the decay rate of non-equilibrium phase fraction. This confirms that it was originating from the percolation effects.

For a proper comparison of the magnetization results with that of resistivity, the magnetization data is also converted to phase fractions. We have used zero field cooled M-H curve at 30~K to perform this conversion.\cite{endnote2} The zero filed cooled state is fully arrested AF-I phase and initial M-H curve behaves like AF. The extrapolation of this AF M-H curve can give the magnetization of AF-I phase at any field. On increasing the field, AF-I phase convert to FM-M phase and at 14~T we have completely FM-M phases. On reducing the field system behaves as FM and this portion of M-H curve can be used to obtain magnetization of FM-M phase at any given field. Now the magnetization (M) can be expressed in terms of contribution from constituent phases and their respective phase fraction
\begin{equation}
M= M_{AF}f_{neq} + M_F(1-f_{neq})\label{eq:M phase fraction}
\end{equation}

Using equation \ref{eq:M phase fraction} the time dependent magnetization data is converted to the time dependent non-equilibrium phase fraction. Being free of percolation effects, the magnetization is a direct probe of phase fractions and so the non-equilibrium phase fraction obtained from magnetization data are more accurate than that from resistivity. The time decay of non-equilibrium phase fraction deduced from the time dependent magnetization data is fitted with the stretched exponential function of equation~\ref{eq:Stretched-Exponential}, which gives the $f_{neq}(0)$  and decay rate ($D$) for various cooling fields. This cooling field dependence of $f_{neq}(0)$  and $D$ is compared with that of resistivity in figure \ref{fig: Decayrate}.
%In figure \ref{fig: Decayrate} (a) we have plotted the cooling field dependence of initial non equilibrium phase fraction ($f(0)_{neq}$) at different temperatures and thermal history.

The $f(0)_{neq}$ calculated from the time dependent resistivity data at 28~K is somewhat off from that of magnetization measurements which suggests that our resistivity to volume conversion is adding some error in the value of $f(0)_{neq}$. But the similar cooling field dependence of $f(0)_{neq}$ (and $D$) of the resistivity and magnetization indicates that the error added in the calculation of $f(0)_{neq}$ does not change their cooling field dependence. The similarity in the field dependence of $f(0)_{neq}$ and $D$ obtained from resistivity and magnetization measurements, even when they have different absolute values is quite surprising. However, this brings out an important aspect that the dynamics in the supercooled (unarrested) and kinetically arrested regimes are so
different that it comes out even if the details of the percolating network is not accurately modelled (as it is not taken care in equation \ref{eq:GEM}),
as long as the resistivity is measured around the percolation threshold, which is evident from figure \ref{fig: Decayrate} (a). In this study, the resistivity
around percolation threshold is chosen by tuning the metallic (FM-M) phase fraction accrued while cooling in different chosen magnetic field range.
Therefore even if shifted, the $f(0)_{neq}$ and $D$ values calculated from resistivity give proper field dependence and are good enough to be used in qualitative analysis. % and so, ignoring the error in resistivity to volume conversion, resistivity $f(0)_{neq}$ and $D$ are used to analyze the temperature, field, and thermal cycle dependence.

To study the effect of temperature and non-equilibrium phase fraction on the dynamics of phase transformation, we have also performed time dependent resistivity measurements at 25~K in cooling run, and at 24~K and 28~K in the heating run. In the heating run, system was cooled from room temperature to 5~K in presence of field and then heated back to temperature of interest, and after temperature stabilization resistivity was recorded as a function of time keeping field constant. These time dependence resistivity results are also converted to volume fraction and then fitted to equation \ref{eq:Stretched-Exponential} to get the characteristic parameters $D$ and $f(0)_{neq}$. In figure \ref{fig: Decayrate} we have compared H-T path dependence of $D$ and $f(0)_{neq}$.

The field variation of $f(0)_{neq}$ is quite small till 2.9~T, and thereafter $f(0)_{neq}$ decays rapidly on increasing the field. This suggest that in approaching the set temperature, depending on the cooling field, two entirely different dynamical mechanisms are determining the initial AF-I phase fraction. Below 2.6~T cooling field, $f(0)_{neq}$ does not exhibit any significant temperature or thermal history dependence, this suggests that phase faction of AF-I and FM-M phases does not change on varying the temperature from 28~K to 25~K, 25~K to 5~K, and heating back to 24~K and then to 28~K. The temperature and thermal history (the difference in cooling and heating cycle) dependence of $f(0)_{neq}$  increases on increasing the field and this difference in $f(0)_{neq}$  between  cooling and heating cycle  becomes nearly constant above 2.9~T cooling filed. This imply that a crossover from one dynamic regime to other happens around 2.9~T.  %This imply that in between 2.6 to 2.9~T cooling field the said two dynamic mechanisms competes, and on increasing the cooling field above 2.6~T the effect of below 2.6~T dynamics diminishes. %At field higher than 3T, a significant fraction of AF-I  phase transform to FM-M phase on the thermal cycling 28K-5K-28K.

Now to understand the dynamical mechanisms mentioned in the above paragraph we recall that in a broad first order transition system, the temperatures related to thermodynamic transition ($T_c$), limit of supercooling ($T^*$), limit of superheating ($T^{**}$), and kinetic arrest ($T_g$) are distributed over a range arising from different regions of sample. Below the transition temperature, PCMAO have FM-M ground state but the AF-I phases may exist in their metastable state.
%Now to understand the dynamical mechanisms mentioned in the above paragraph we recall that in a broad first order transition system, the values of transition temperature, temperature of metastabilty, and temperature of kinetic arrest for different phases are distributed over a temperature range. Below the transition temperature, PCMAO have FM-M ground state but the AF-I phases may exist in their metastable state.
%Now to understand the dynamical mechanisms mentioned in the above paragraph we note that below the transition temperature, PCMAO have FM-M ground state while the AF-I phases are present in their metastable state.
The AF-I phases present in their metastable state can transform to the FM-M ground state and the rate of transformation is determined by two different dynamical process having different time constants. Of these, the first process is thermodynamic which gives the time constant of formation of FM-M phase critical nuclei in the metastable AF-I phase ($\tau_1$) and this time constant (and so the thermodynamically allowed rate of transformation) is completely determined by the energy of nucleation. The second process is the growth of FM-M phase critical nuclei to complete the thermodynamically allowed transformation. This process depends on the kinetics of the constituent atoms and it slows down exponentially on approaching the temperature-field ($T_g$-$H_g$) of kinetic arrest. Far above $T_g$-$H_g$ (un-arrested regime) the time taken in the growth of critical nuclei ($\tau_2$) is small\cite{Sharma} and so the rate of transformation is primarily determined by the energy of nucleation. On approaching $T_g$-$H_g$ the time taken in the growth of critical nuclei progressively increases and at some temperature it crossover the time in which the transformation is thermodynamically allowed. Below this temperature-field value (arrested regime) the rate of transformation is mainly determined by time taken in the growth of critical nuclei.

In figure \ref{fig: Decayrate}(a), the change in cooling field dependence of $f(0)_{neq}$ on lowering the field below 2.9~T suggests a crossover from  un-arrested to arrested dynamics. Around 2.9~T, $\tau_1$ and $\tau_2$ are comparable and $\tau_2$ increases exponentially on decreasing the cooling field. Below 2.6~T, $f(0)_{neq}$ has negligible field, temperature and thermal history dependence. This indicates that in this field region $\tau_2$ is huge in comparison to $\tau_1$ and so the system is in arrested regime.  Above 2.9~T, the system moves outside the region of kinetic arrest  and  in this field region the rate of transformation is mainly determined by the energy of nucleation. In this un-arrested regime, a fraction of AF-I phases transforms to FM-M state during thermal
cycling (28~K - 5~K - 28~K) and this transformation gives a thermal history dependence in $f(0)_{neq}$. We also note that in 28~K heating run, the change in
the cooling field dependence of $f(0)_{neq}$ happens at a lower field in comparison to the cooling run. At the same temperature, heating run $f(0)_{neq}$ is
small compared to that of cooling run and this suggests that on lowering $f(0)_{neq}$ the crossover from
un-arrested to arrested dynamics happens at a lower field.

Figure \ref{fig: Decayrate}(b) show the variation of decay rate ($D$) with cooling field at different temperatures and also with thermal history. At 28~K, the cooling field dependence of decay rates obtained from resistivity and magnetization measurements are qualitative similar. The decay rate is quite small at low cooling fields. This is because at low cooling field we are in the region of kinetic arrest and so the time taken in the growth of critical nuclei ($\tau_2$, which is huge) is dominating the transformation rate and this results in a smaller decay rate. On increasing the field, AF-I phases starts moving away from the region of kinetic arrest and this reduces the $\tau_2$ which results in a weak growth in the decay rate. Above 2.9~T cooling field, the decay rate exhibit a faster growth on increasing the cooling field. This is because above 2.9~T, the system is away from the region of kinetic arrest and now the rate of transformation is decided by the energy of nucleation. The energy of nucleation depends on the difference in the free energies of the supercooled metastable and the ground state and it decreases as this difference increases.\cite{Callister}  In PCMAO the metastable state is AF-I and the ground state is FM-M; the increase in field decreases the free energy of FM-M state which in turn enhances the free energy difference and reduces the energy of nucleation. Therefore decay rate show a faster growth rate  on increasing the field above 2.9~T. A comparison of 28~K cooling and heating cycle decay rate suggest that the de-arrest happens at a lower field range on decreasing the $f(0)_{neq}$. This is in agreement with the observation made in the previous paragraph.

At 28~K cooling run, on increasing the cooling field above 3.25~T, the decay rate slowly decreases on increasing the field. The increase in cooling field lowers the nucleation energy. This lowering of nucleation energy enhances the allowed rate of transformation but it will increase the decay rate of AF-I phase fraction only till there is a certain number of supercooled metastable AF-I phases left. This is because the amount of volume transferred from AF-I to FM-M state in a certain time period depends on the allowed rate of transformation and on the number of available metastable AF-I phases that can transform. In this case as the field is increased above 3.25T, $f(0)_{neq}$ decreases below 0.78 (of resistivity) and this decrease in available metastable AF-I phase fraction starts slowing down the decay rate $D$. When the temperature is lowered below 28~K the phase fraction of initial FM-M phase (1-$f(0)_{neq}$) increases. The enhancement in the number of FM-M phases increases the nucleation centers which in turn decreases the energy of nucleation. Because of this, far from the kinetic arrest region, the decay rate of 28~K heating is higher than the decay rate of 28~K cooling. Similar is the case of 24~K heating and 25~K cooling. In a particular run (cooling or heating), the decay rate decreases on decreasing the temperature of measurement. This indicates that on temperature lowering the decay in nucleation energy is slower than the decay in thermal energy.  At all temperatures, the decay rates start decreasing as the initial AF-I phase fraction goes below around 0.78 (of resistivity).

\section{Conclusion}
In conclusion, our results and discussion on PCMAO show that the dynamical effects observed in the phase separated state can be effectively controlled by
the cooling field, temperature of measurement, and phase fraction of metastable AF-I phases. These dynamical effects originate from the transformation of
metastable AF-I phases to FM-M ground state and the observed rate of transformation depends on two competing and drastically different dynamic mechanisms, the first is the formation of FM-M critical nuclei in the AF-I metastable phase and the second is growth of this critical nuclei to complete the phase transformation, with different time constants. The observed rate of transformation exhibits two distinct dynamics regimes, the arrested regime where time taken in growth of critical nuclei is huge compared to the time taken in formation of critical nuclei and the un-arrested regime where the time taken in the growth of critical nuclei is negligible
compared to the time taken in formation of critical nuclei. The dynamics of the phase transformation transverse from arrested to un-arrested regime on increasing the temperature and cooling field and at a given temperature the crossover from arrested to un-arrested dynamics happens at lower field on decreasing the metastable phase fraction. In the un-arrested regime the rate of AF-I to FM-M transformation depends on the energy of nucleation and the
available phase fraction of metastable AF-I phases.

%We have performed detailed time dependent resistivity and magnetization measurements on PCMAO under different field, temperature, and thermal cycle. Our results
%and discussions show that in the

%the results in a smalle and suggest that of the two dynamic mechanisms i.e the thermodynamically allowed transformation of AF-I to FM-M phases and the glass like arrest of transformation kinetics, the later is dominating and is dictating the decay rate. This is in agreement with the conclusions drawn from the field dependence of  $f(0)_{Neq}$. On increasing the field, the temperature of kinetic arrest shifts towards lower temperature and we move away from the kinetic arrest band. In this de-arrested scenario, the time required for rearranging the atoms becomes progressively small and so the dynamics will be governed by the time required for thermodynamic transformation which in turn is determined by the energy of nucleation.


\begin{thebibliography}{10}
\bibitem{endnote1} The magnetization of PCMAO at 5~K in  10~T cooling field is 3.4 $\mu B$/f.u and the saturation magnetization of PCMAO is around 3.5 $\mu B$/f.u. So the assumption that the 10~T cooled state is FM-M adds a maximum error of about 3\%.
\bibitem{endnote2} We have used 30~K M-H loop in calculating the magnetization of FM-M and AF-I phases. This gives a 4\% lower magnetization
value for FM-M phase at 2~T, this error decreases on increasing the field and at 4~T it is 1\%. The error in the value of AF-I phase magnetization is negligible.
\end{thebibliography}
\end{document}